\documentclass[letter]{aa}  

\usepackage{graphicx}
\usepackage{txfonts}
\begin{document}

   \title{A massive mess:\\When a large dwarf and a Milky Way-like galaxy merge}
   \titlerunning{A massive mess}
   \author{Helmer H. Koppelman \and Roy O.Y. Bos
          \and
          Amina Helmi
          }

   \institute{Kapteyn Astronomical Institute, University of Groningen, Landleven 12, 9747 AD Groningen, The Netherlands\\
              \email{koppelman@astro.rug.nl}
}
   \date{Received: ; accepted: }

  \abstract
   {}
   {
   Circa 10 billion years ago the Milky Way merged with a massive satellite, Gaia-Enceladus. To gain insight into the properties of its debris we analyse in detail the suite of simulations from \cite{Villalobos2008}, which includes an experiment that produces a good match to the kinematics of nearby halo stars inferred from {\it Gaia} data.
   }
   {
   We compare the kinematic distributions of stellar particles in the simulations and study the distribution of debris in orbital angular momentum, eccentricity and energy, and its relation to the mass-loss history of the simulated satellite.
   }
   { We confirm that Gaia-Enceladus probably fell in on a retrograde,
     30$^\circ$ inclination orbit. We find that while 75\% of the
     debris in our preferred simulation has large eccentricity
     ($> 0.8$), roughly 9\% has eccentricity smaller than 0.6. Star
     particles lost early have large retrograde motions, and a subset
     of these have low eccentricity. Such stars would be expected to
     have lower metallicities as they stem from the outskirts of the
     satellite, and hence naively they could be confused with debris
     associated with a separate system. These considerations seem to
     apply to some of the stars from the postulated Sequoia galaxy. 
   }
   { When a massive discy galaxy merges, it leaves behind debris with
     a complex phase-space structure, a large range of orbital
     properties, and a range of chemical abundances. Observationally, this
     results in substructures with very different properties, which can
     be misinterpreted as implying independent progeny. Detailed
     chemical abundances of large samples of stars and tailored
     hydrodynamical simulations are critical to resolving such
     conundrums.}


   \maketitle
%

\section{Introduction}

The ultimate goal of galactic archaeology is to determine the series of events that have led to the formation of the Milky Way. The arrival of the full phase-space dataset of {\it Gaia} \citep{GaiaCollaboration2016TheMission, GaiaCollaboration2018brown, Katz2019GaiaVelocities} has resulted in many insights that have given a boost to this field. One of these insights is that the local stellar halo formed predominantly through a merger with a massive dwarf galaxy named Gaia-Enceladus \citep[][or Gaia-Sausage, \citealt{Belokurov2018Co-formationHalo}]{Helmi2018}. 

It remains unclear, however, which other objects on retrograde orbits have contributed debris to the local stellar halo. Although several smaller retrograde structures have been found that appear to be chemically and dynamically different from Gaia-Enceladus \citep[e.g.][]{Koppelman2018, Mackereth2019TheSimulations, Matsuno2019, Myeong2019EvidenceHalo}, their origin and linkage are not always clear. For example,  
\cite{Myeong2019EvidenceHalo} advocate that several of these structures originate in a single (moderately) massive dwarf galaxy: Sequoia. However, \cite{Koppelman2019} based on their chemistry and orbital properties, suggest that there must be at least two progenitors of the structures. 

The debris of a low-mass satellite orbiting in a static potential
phase-mixes at roughly constant mean orbital energy \citep{Helmi1999a}. This
makes the integrals of motion powerful tools to identify halo
structures and determine accretion history \citep{helmi2000,
  McMillan2008}. For massive satellites, this picture becomes muddled
because of dynamical friction and the tidal interaction between the
satellite and host \citep[see
e.g.][]{Jean-Baptiste2017OnTale}. Dynamical friction is relevant for
high mass-ratios between the satellite and its host. It affects the
mean orbit of a satellite and makes it sink to the centre of the host
\citep{Quinn1986SINKINGSYSTEMS}. Since a satellite is stripped of its
mass mostly in discrete events at its orbital pericentre (and this
evolves because of dynamical friction), the result is a complex energy
distribution of the debris. For example, the material from the
satellite's outskirts is lost early and hence has high orbital
energy. On the other hand, its core sinks to the centre becoming more
bound \citep[][]{Tormen1998SurvivalHaloes,
  VanDenBosch1999SUBSTRUCTUREFRICTION}. Furthermore, massive
satellites with disc-like morphology produce very complex tidal
features \citep{Quinn1984OnGalaxies} which must give rise to an intricate structure of
the remaining debris. These considerations are pertinent for
understanding the left-overs from a system like Gaia-Enceladus.

\begin{figure*}
    \centering
    \includegraphics[width=\hsize]{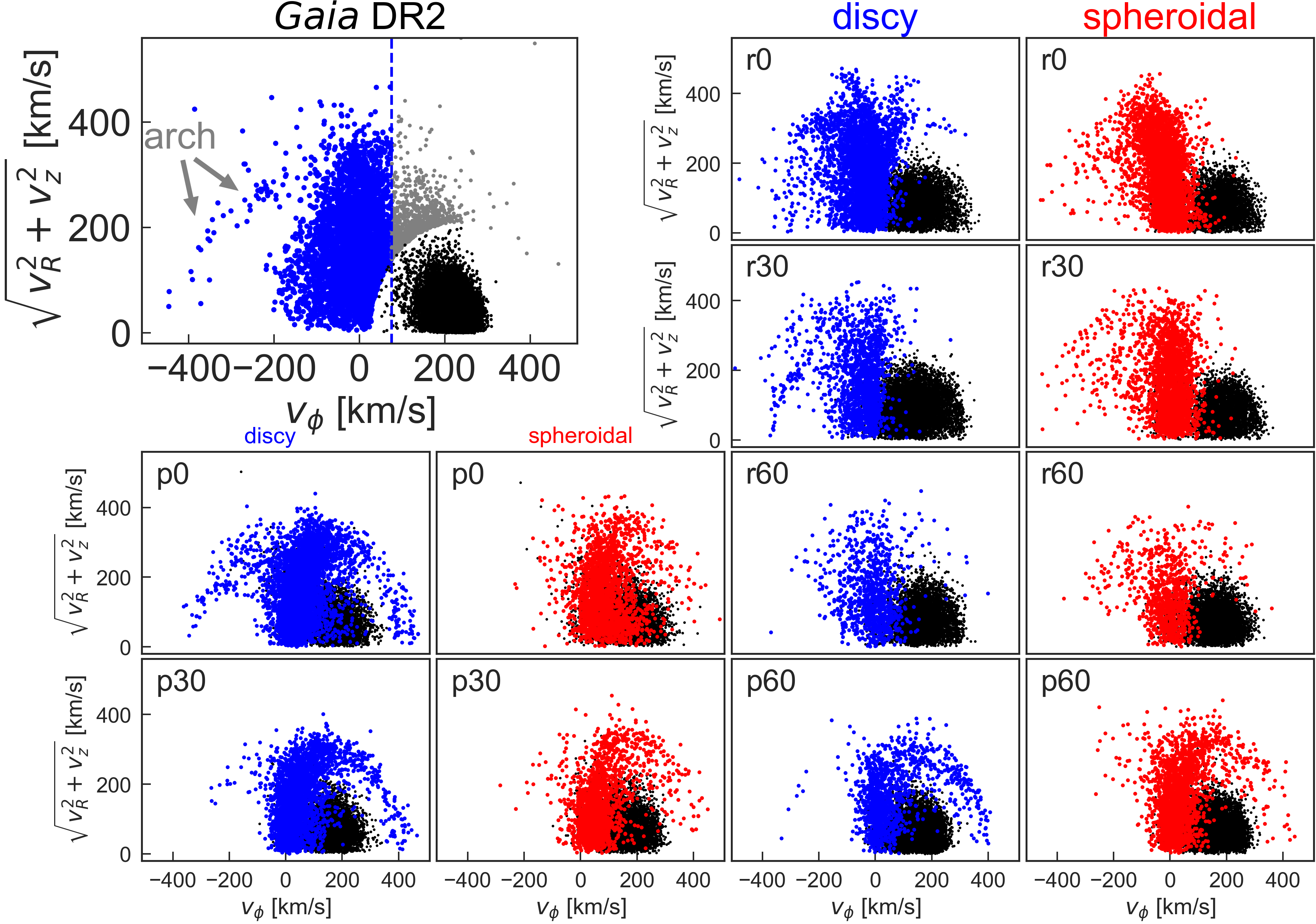}
    \caption{Velocity distribution of the local $(<1~{\rm kpc})$ stellar halo using {\it Gaia} data (large panel) and in the \cite{Villalobos2008} simulations, where the insets indicate the orbital inclination $(0^\circ, 30^\circ~{\rm or}~60^\circ)$ and whether the merger was prograde or retrograde. Stellar particles from the discy (spherical) satellites are shown in blue (red). These stars' distinct velocity distributions betray the progenitor's properties and the merger geometry.}
    \label{fig:velocity}
\end{figure*}

Further complexity is expected from the chemical perspective, since dwarf galaxies also display chemical gradients \citep[e.g.][]{Kirby2011MULTI-ELEMENT, Ho2015MetallicityMetals}. These gradients are typically negative with radius because the star-formation-rate in the centre of galaxies is more intense due to higher gas density. Moreover, star-formation in the outskirts may be quenched first because this is where gas is stripped more easily. This means that merger debris should probably also reveal chemical gradients. Such a gradient has been found for Sagittarius \citep{Ibata1994ASagittarius} as its streams have older stellar populations \citep{Bellazzini2006DetectionStream}, and are more metal-poor by $\sim 0.7~{\rm dex}$ than the core \citep[see][and \citealt{Hayes2020MetallicityAPOGEE}, for more references]{Dohm-Palmer2001MAPPINGBODY, Martinez-Delgado2004TracingGalaxy, Chou2007AStream}. 

Motivated by the above considerations, in this Letter we use numerical simulations to study the dynamical gradients that naturally occur in mergers of massive satellites. The results are particularly important to interpret the debris of Gaia-Enceladus and its relation to the other substructures discovered in the halo, such as the Sequoia galaxy.
\section{Methods}\label{sec:methods}

We analyse the merger simulations of \citet{Villalobos2008, Villalobos2009SimulationsDiscs} designed to study the establishment of a thick disc through heating of a thin disc by a merger. The simulation suite comprises a set of 1:10 and 1:5 mass-ratio mergers of varying orbital inclination and of both spheroidal and discy satellites. After the merger, part of the host's disc is heated to a plausible thick disc. At the same time, a significant fraction of the original disc $(15-25\%)$ remains thin and cold. We focus on these simulations because \cite{Helmi2018} noted that the kinematic distribution of the satellite in one of the experiments (denoted here as {\tt r30}) matches very well that of nearby Gaia-Enceladus stars.

Since the host galaxy in the simulations is smaller than the Milky Way,  and to facilitate comparison to observations, we scale the velocities such that the rotational velocity of the simulated disc at the solar position $v_{\phi,\odot,{\rm sim}}$, corresponds to that measured for the Milky Way's thick disc \citep[$v_{\phi,\odot} = 170~{\rm km/s}$, according to][]{Morrison1990}. Following \cite{Villalobos2009SimulationsDiscs} we place the Sun at $2.4 R_D$, and we scale the velocities by the factor $v_{\phi,\odot}/v_{\phi,\odot,{\rm sim}}$. For example for the experiment {\tt r30}, $R_D \approx 2.16~{\rm kpc}$ \citep[see Fig.~14 of][]{Villalobos2008}, $v_{\phi,\odot,{\rm sim}} \sim$~132 km/s, and the scaling factor is $\sim 1.3$. 
 
The top-left panel of Fig.~\ref{fig:velocity} shows the velocity distribution of a sample of halo stars in {\it Gaia} DR2 (see Appendix~\ref{app:RVS} for details). Halo stars with $v_\phi<75~{\rm km/s}$ (to the left of the dashed line) are highlighted with blue dots. For illustrative purposes, we indicate the location of the thin disc with a random subset of $50~000$ stars (black dots). The remaining panels in Fig.~\ref{fig:velocity} show the velocity distributions for all the 1:5 mass-ratio simulations, with the stellar particles from the host in black and those from the satellite in blue for discy and red for spheroidal progenitors. These particles are located inside a volume of 2.5 kpc radius centred on the `equivalent' solar position at the end of the simulations, i.e. 4 Gyr after infall. Because of particle resolution limitations, this volume is relatively large compared to that used for the data, especially because it has not been scaled.

\begin{figure}
    \centering
    \includegraphics[width=\hsize]{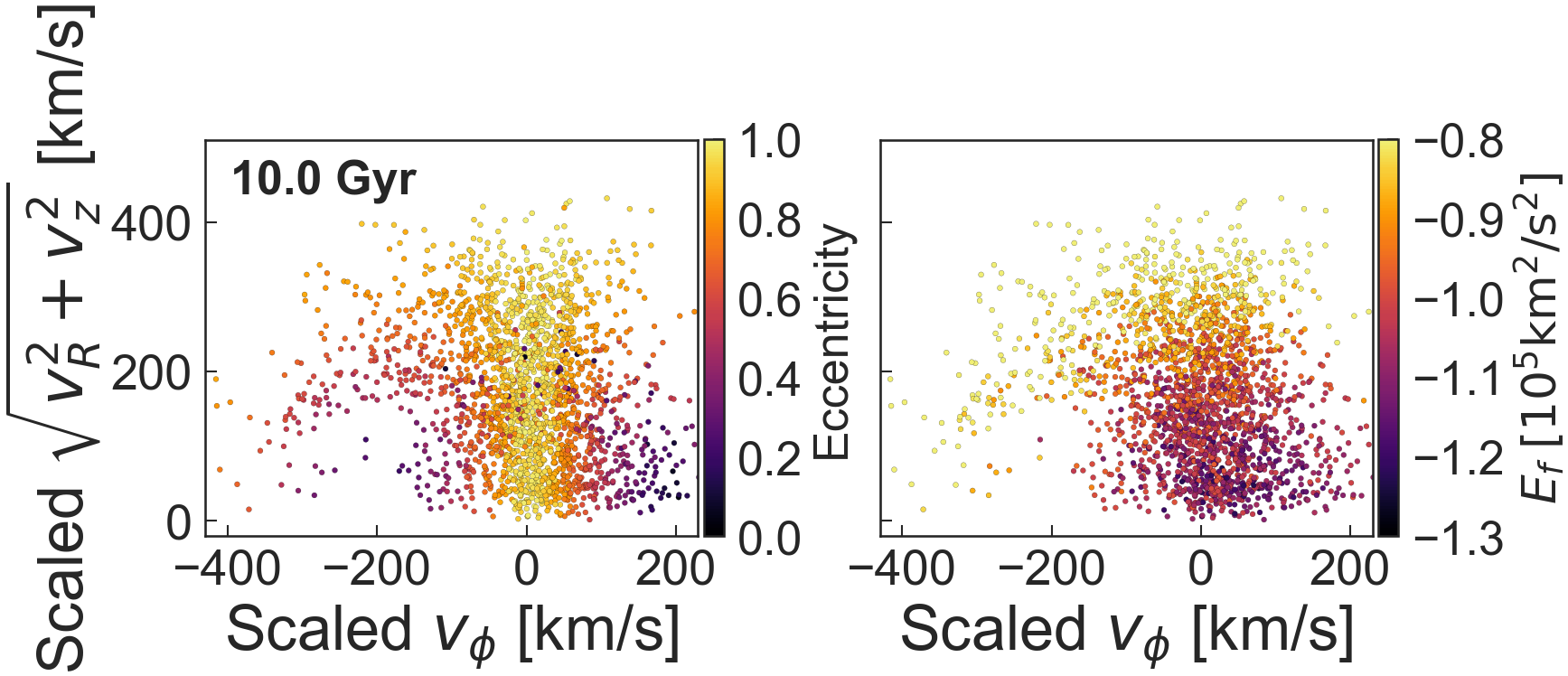}
    \caption{Velocity distribution for the discy {\tt r30} simulation 10 Gyr after infall. The stellar particles are coloured according to their final eccentricity (left) and energy ($E_f$, right). The extended `arch'-like structure (with very negative $v_\phi$) is persistent in time, see for comparison Fig.~\ref{fig:velocity}.}
    \label{fig:velcol}
\end{figure}

As reported in \cite{Helmi2018}, the arch seen in the {\it Gaia} data
and shown in the top-left panel of Fig.~\ref{fig:velocity} is well reproduced in the discy
simulation with the inset {\tt r30}, suggesting that the progenitor of
Gaia-Enceladus was discy and merged on a retrograde orbit of
$\sim30^\circ$ inclination. However, because this simulation was only
run for 4 Gyr, to establish the robustness of this conclusion and the arch's origin, we have integrated the simulation for a
total of 10 Gyr, which is 
approximately the time of the merger with Gaia-Enceladus.
\section{Results}\label{sec:results}

\begin{figure}
    \centering
    \includegraphics[width=\hsize]{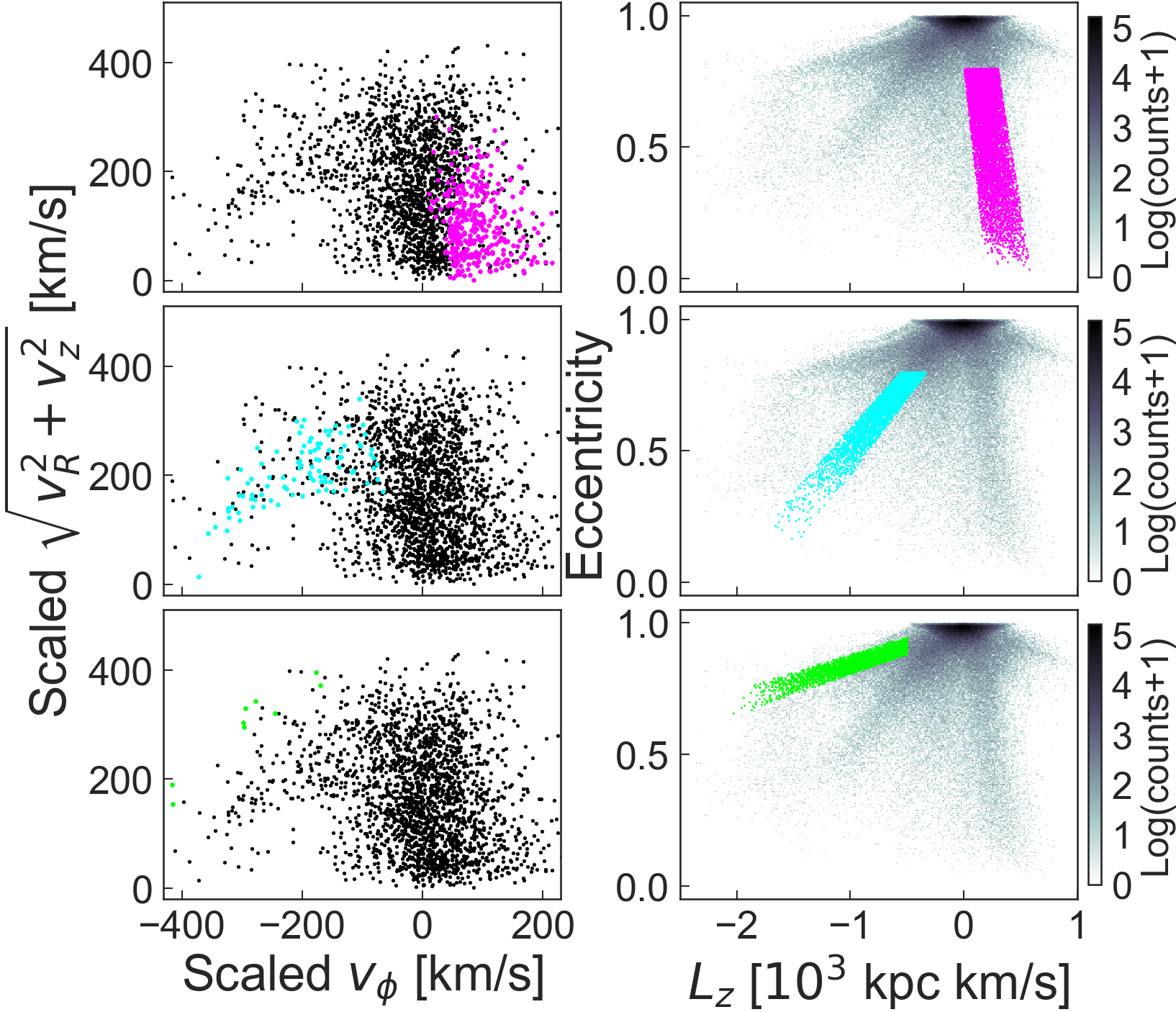}
    \caption{Velocity distribution in a solar volume (left, same as in Fig.~\ref{fig:velcol}) and 
distribution of $e$-$L_z$ of all satellite stellar particles (right) both at $t=10$~Gyr. The colours  
indicate the location of various selections made in $e$-$L_z$, following the three clear ridges identifiable
in this space.}
    \label{fig:lzecc}
\end{figure}

\subsection{Origin of the velocity arch}

Figure~\ref{fig:velcol} shows the velocity distribution in a representative solar volume of the {\tt r30} simulation 10 Gyr after infall. Although the shape of the distribution depends slightly on the azimuthal location of the volume (because the remnant thick disc in the simulations is somewhat triaxial), the arch structure in velocity space is seen to persist in time. This implies that the conclusions that are drawn on the basis of Fig.~\ref{fig:velocity} still stand. 

To characterise the arch we have coloured the stars in Fig.~\ref{fig:velcol} by their final eccentricity $e$ (left panel) and orbital energy $E_f$ (right panel), see Appendix~\ref{app:orbpar} for details. Note the large range of values of both quantities and how they appear to correlate with the velocities, although not perfectly. Since its discovery, Gaia-Enceladus has been conflated with stars on very eccentric $(e \gtrsim 0.8)$ orbits \citep[e.g.][]{Belokurov2018Co-formationHalo, Mackereth2019TheSimulations}. Even though this might be true for the bulk of the stars, we find that this is not necessarily true for all of its debris. Roughly $75\%$ of the stars ends up on orbits of $e>0.8$, while $\sim 9$\% have eccentricities $e<0.6$. 

Figure~\ref{fig:lzecc} shows the link between the local structure in
velocity space (left) and the orbital properties of all of the debris
in $e$-$L_z$ space. The latter is a function of velocity
($L_z = R v_\phi$) and is an integral of motion often used to identify
Gaia-Enceladus' debris \citep[e.g.][]{Helmi2018, Matsuno2019,
  Massari2019OriginWay}. The satellite's debris displays a large
amount of substructure as shown in the panels on the right, which is
clearly linked to that present in velocity space in the solar
volume. For example, the arch in velocity space appears to be the local
manifestation of the large retrograde structures marked in the middle and bottom rows of Fig.~\ref{fig:lzecc}. On the other hand, the top row of this figure highlights that some stars have relatively prograde angular momentum despite the initially retrograde orbit of the satellite.

\begin{figure*}
    \centering
    \includegraphics[width=\hsize]{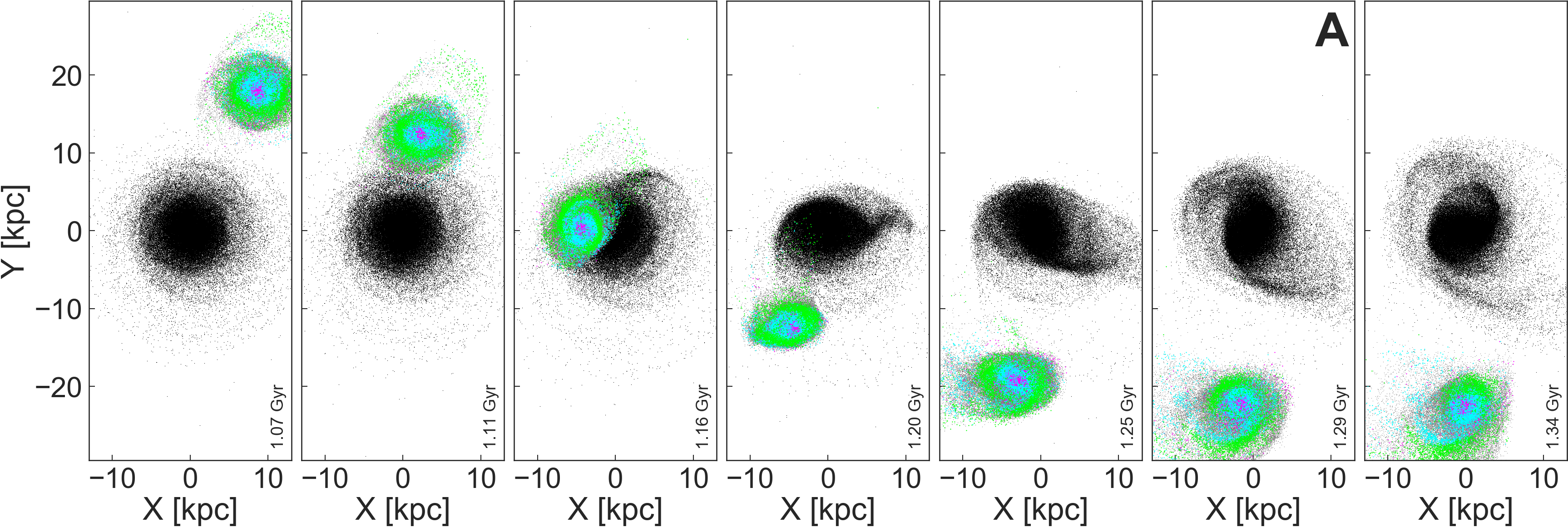}
    \includegraphics[width=\hsize]{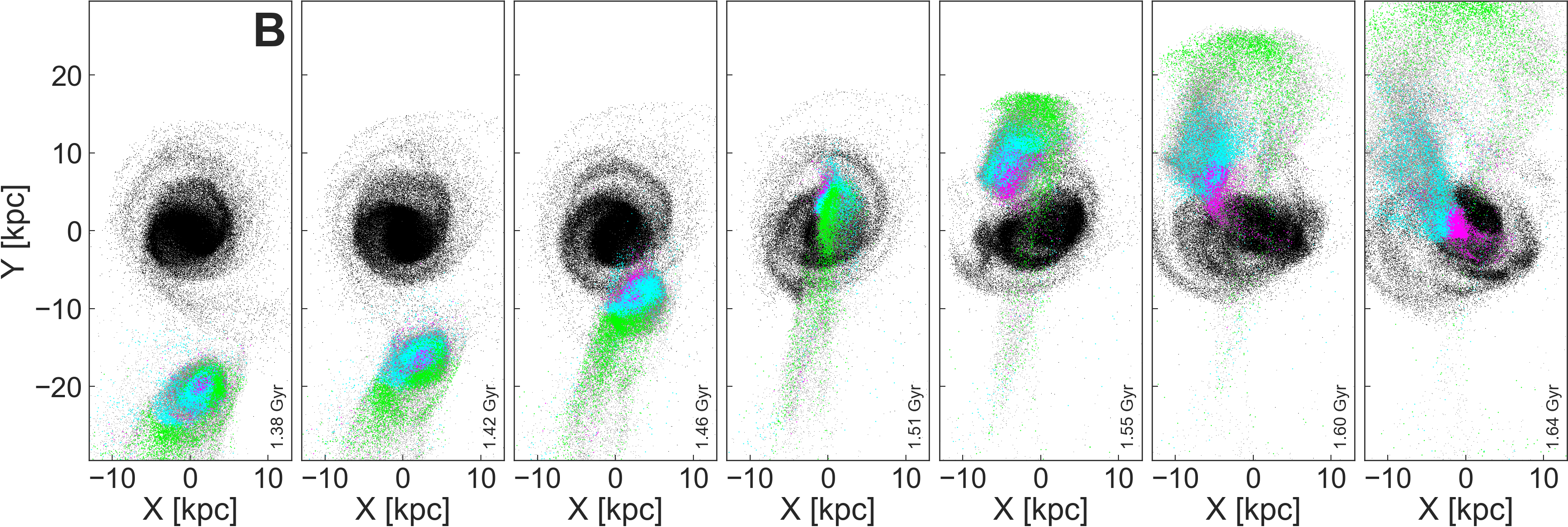}
    \includegraphics[width=0.6\hsize]{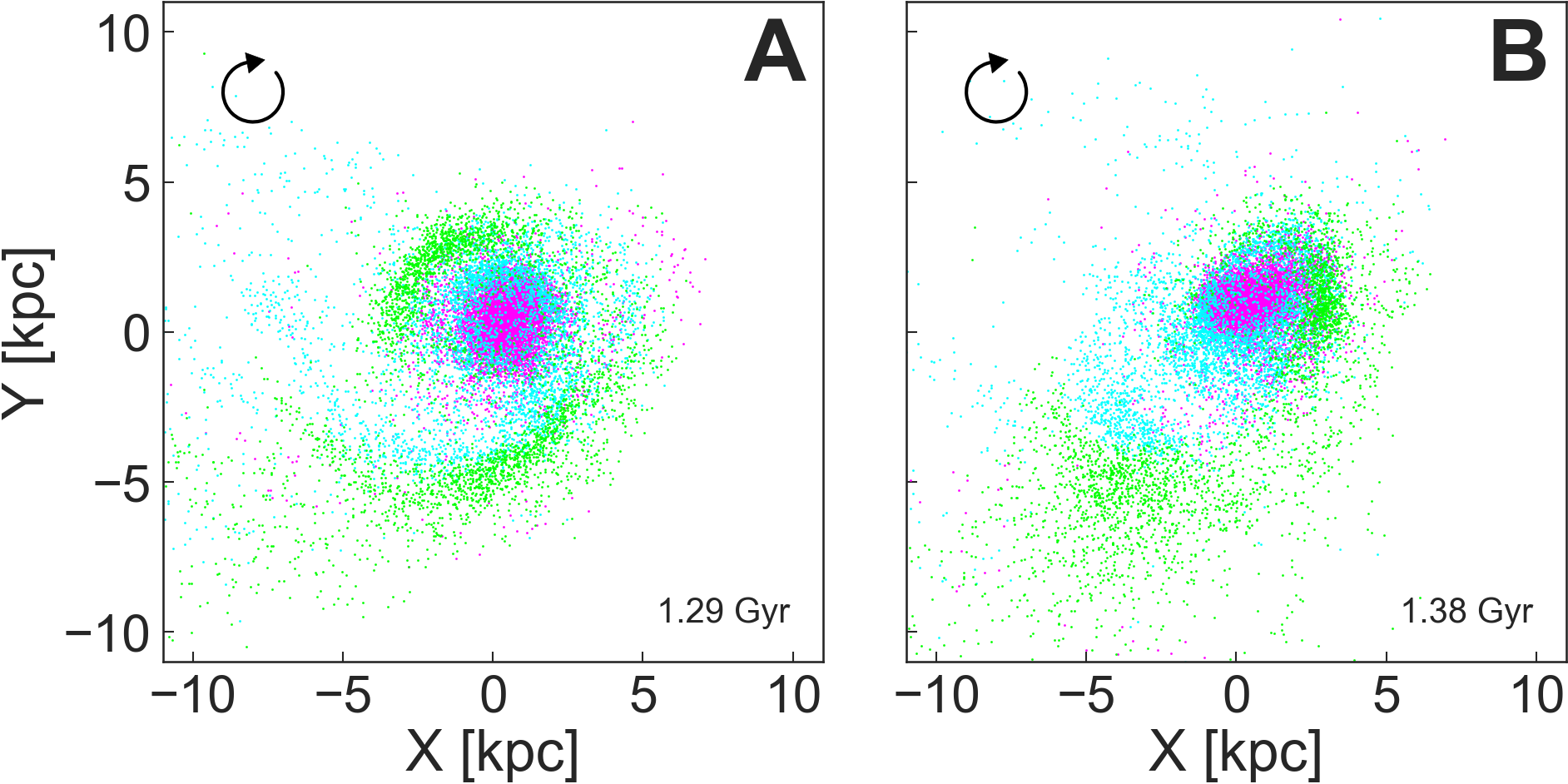}
    \caption{Evolution of the satellite around the time of the second pericentric passage, which takes place at $t\approx1.5$~Gyr. The colours trace the three structures identified in the final orbital distribution of the debris, see Fig.~\ref{fig:lzecc}, and evidence the complexity of the mass loss process. This is clearly seen in the bottom two panels which zoom-in onto the satellite at two specific times, and where its spin is indicated with a curled arrow. We encourage the reader to also inpect the animation online.}
    \label{fig:massloss}
\end{figure*}

We investigate the origin of the structures in $e$-$L_z$ space in Fig.~\ref{fig:massloss}, where we plot the spatial evolution of the satellite between $1-1.7$~Gyr after infall. In the first panel, at $t \sim 1$ Gyr, most of the satellite's dark matter has been lost but the stellar component remains largely bound. A significant fraction of the stars is lost during/shortly after the second pericentric passage which takes place at $t\sim 1.5$~Gyr \citep{Villalobos2008}. The core (traced largely by the magenta points) continues to spiral inwards for two to three more passages and then fully dissolves. During this process, it couples more strongly to the disc and becomes more prograde (as we saw in the top right panel of Fig.~\ref{fig:lzecc}). To appreciate the high complexity of the mass loss process, an animation of Figure~\ref{fig:massloss} is available online\footnote{\url{https://www.astro.rug.nl/~ahelmi/MassiveMess}}.

When cold rotationally supported disc galaxies merge they produce
complex tidal tails \citep{Toomre1972GalaxyTails,
  Eneev1973TidalGalaxies, Quinn1984OnGalaxies,
  Barnes1988EncountersGalaxies}. These tidal tails can be quite
different from those originating in spherical, dispersion supported
systems. This is also what we find in our simulations as shown in
Fig.~\ref{fig:massloss}. The bottom row of this figure zooms in on the
discy satellite and reveals that some of the stars in green, which end
up on high eccentricity, very retrograde orbits (i.e.  the ridge in
the bottom panel of Fig.~\ref{fig:lzecc}) are lost first.  These are followed by some 
stars in cyan with lower final eccentricity and
slightly less retrograde motions since they are part of the ridge in the middle
panel of Fig.~\ref{fig:lzecc}. We also see that there is some
alternation in green and cyan stars amongst the material that becomes
unbound slightly later. Since a discy satellite has internal as well
as orbital angular momentum, mass loss depends on the exact
configuration of the merger (e.g. inclination, spin) and not only on
the internal binding energy of the stars as in the case of a
spheroidal system. A difference in eccentricity then arises from these
additional degrees of freedom. For a given internal binding
energy stars have a range of internal angular momenta, which translates in
debris with high- and low orbital angular momentum, resulting in
more circular or more radial final orbits, respectively.

On top of these effects, dynamical friction
causes the satellite's core to spiral inwards. This enhances the prominence of the only tidal arm
apparent in Fig.~\ref{fig:massloss},  which trails behind the core.


\subsection{Some implications}

We have just shown that the stars in the velocity arch constitute
debris lost early on in the merger. This probably has also
implications on the chemical abundances of its stars since dwarf
galaxies are known to depict metallicity gradients. For
$M_\star\sim 10^{9.6}~{\rm M}_\odot$ the gradients have an amplitude
$\sim -0.064~{\rm dex/kpc}$ according to
\citet{Ho2015MetallicityMetals}. Thus if Gaia-Enceladus had a physical
extent of 5 kpc, this could imply a metallicity difference of
$0.32$~dex between its centre and outskirts.

\cite{Myeong2019EvidenceHalo} studied some of the stars in the
velocity arch and argued that these, together with other retrograde
stars, were part of a different accreted galaxy which they named
Sequoia. These authors have shown that the stars are more metal-poor
on average than the debris of Gaia-Enceladus by $\sim 0.30$ dex, which
suggests they formed in a ten times smaller object
\citep{Matsuno2019}. However, an alternative explanation supported by
our simulations is that (some of) these stars stem from the outskirts of
Gaia-Enceladus, and their lower metallicities could be explained as
being due to internal gradients in Gaia-Enceladus. Also the lower
values of [Mg/Fe] and [Ca/Fe] at similar [Fe/H] in comparison to
Gaia-Enceladus reported by \cite{Matsuno2019} could be due to 
the typically lower star formation rates found in the outskirts of
sizeable galaxies.

\begin{figure}
    \centering
    \includegraphics[width=\hsize]{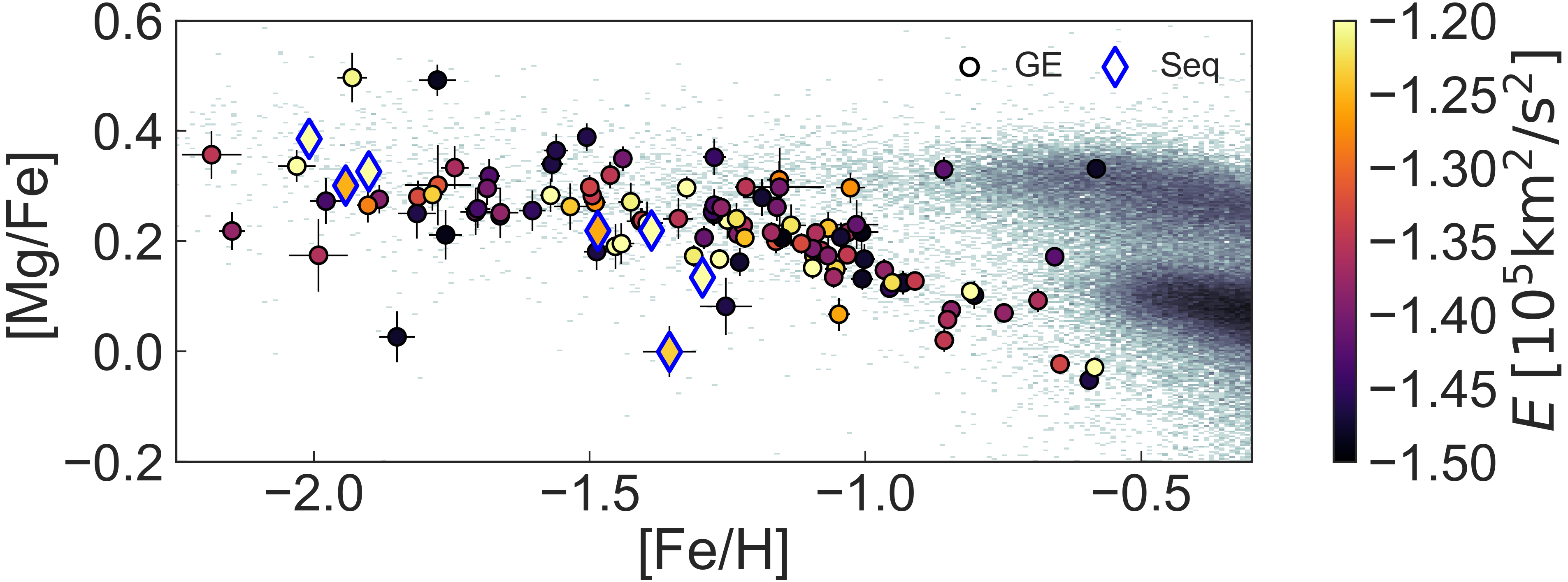}
    \caption{Chemical abundances for stars in Gaia-Enceladus and Sequoia, identified using selection criteria from \cite{Koppelman2019}. The stars are coloured by their orbital energy calculated in the \cite{Mcmillan2017} potential implemented in {\tt AGAMA}. The background shows the location of the disc-stars in this diagram with a 2D-histogram. The histogram is coloured by the logarithm of the stars per bin.}
    \label{fig:APOGEE}
\end{figure}

This interpretation is tentatively supported by Fig.~\ref{fig:APOGEE} where we plot [Mg/Fe] vs [Fe/H] for members of Gaia-Enceladus and Sequoia, coloured by their orbital energy $E$. The chemical abundances are from a cross-match of {\it Gaia} DR2 with APOGEE DR16 \citep{Ahumada2020TheSpectra}, using the membership criteria described in \cite{Koppelman2019} (see the Appendix for more details). This figure shows that most Sequoia stars are accompanied by a star from Gaia-Enceladus that has very similar orbital energy and chemical abundance, making it difficult to argue on the basis of this data that the stars have a different origin.

\section{Conclusions}

We have analysed simulations of the merger between a Milky Way-like galaxy and a massive satellite. On the basis of comparisons to the kinematics of stars in the nearby halo, we confirm the conclusions by \cite{Helmi2018} that the most likely progenitor of Gaia-Enceladus was a discy dwarf galaxy that fell in on a retrograde orbit of $\sim 30^{\rm o}$ inclination. 

Because of the relative mass of the dwarf galaxy (which implies it was
subject to dynamical friction), and its initial configuration (a cold,
rotationally supported disc), its debris depicts a rather intricate
morphology, quite different from that usually associated to small,
spherical, non-rotating satellites. Star particles are found to have a
broad range of orbital eccentricities, with 75\% having $e > 0.8$, and
9\%  $e < 0.6$. We also find that even though the initial orbital motion
of the satellite is very retrograde, some of its debris ends up on
prograde orbits. Low eccentricity and prograde stars originate in the
core of the satellite which has sunk in via dynamical friction. We may conclude that a
large amount of information about the merger remains encoded in the
phase-space structure of the debris.

Because large galaxies have chemical gradients, different portions of the debris may reveal different chemistry. This complicates the interpretation, and can possibly mimic what might be expected for debris from an independent system. For example, some of the stars in the arch (which according to our simulations must have been lost early) are dynamically similar to those that have been associated with Sequoia. On the other hand, the abundances of Sequoia stars do not appear to differ much from highly energetic/less bound
(and hence potentially lost early) stars from Gaia-Enceladus.

Taken at face value, the simulation would predict that the core of the Gaia-Enceladus dwarf should be found on a mildly prograde, lower eccentricity orbit. However, this prediction depends strongly on whether the configuration of the merger is matched in detail by the simulations. In particular, the simulated host galaxy has a fixed size, which is unrealistic for the Milky Way, which in the last 10 Gyr has grown a very significant cold disc. Simulations with varying orientation of the discs' angular momentum vectors, and including gas and star formation physics are necessary to fully exploit the data that is currently available, and that which will be collected in the context of upcoming spectroscopic surveys like WEAVE \citep{Dalton2012WEAVE:Telescope}, 4MOST \citep{deJong20124MOST:Telescope}, SDSS-V \cite{Kollmeier2017SDSS-V:Spectroscopy}, and DESI \citep{DESICollaboration2016TheDesign}. Such a combined approach seems to be necessary as a massive satellite can give rise to substructures that appear to be both chemically and dynamically distinct. Massive mergers are messy.

\begin{acknowledgements}
      This work has been financially supported from a Vici grant and a Spinoza Prize from NWO. We have made use of data from the European Space Agency (ESA) mission Gaia (\url{http://www.cosmos.esa.int/gaia}), processed by the Gaia Data Processing and Analysis Consortium (DPAC, \url{http://www.cosmos.esa.int/web/gaia/dpac/consortium}). Funding for the DPAC has been provided by national institutions, in particular the institutions participating in the Gaia Multilateral Agreement. In the analysis, the following software packages have been used: {\tt vaex} \citep{Breddels2018}, {\tt numpy} \citep{VanDerWalt2011TheComputation}, {\tt matplotlib} \citep{Hunter2007Matplotlib:Environment}, {\tt jupyter notebook} \citep{Kluyver2016JupyterWorkflows}, {\tt pyGadgetReader} \citep{Thompson2014PyGadgetReader:Python}.
\end{acknowledgements}

\bibliographystyle{aa} 
\bibliography{aanda} 

\begin{thebibliography}{47}
\expandafter\ifx\csname natexlab\endcsname\relax\def\natexlab#1{#1}\fi

\bibitem[{Ahumada {et~al.}(2020)Ahumada, Prieto, Almeida, Anders, Anderson,
  Andrews, Anguiano, Arcodia, Armengaud, Aubert, Avila, Avila-Reese, Badenes,
  Balland, Barger, Barrera-Ballesteros, Basu, Bautista, Beaton, Beers,
  Benavides, Bender, Bernardi, Bershady, Beutler, Bidin, Bird, Bizyaev, Blanc,
  Blanton, Boquien, Borissova, Bovy, Brandt, Brinkmann, Brownstein, Bundy,
  Bureau, Burgasser, Burtin, Cano-Diaz, Capasso, Cappellari, Carrera,
  Chabanier, Chaplin, Chapman, Cherinka, Chiappini, Choi, Chojnowski, Chung,
  Clerc, Coffey, Comerford, Comparat, da~Costa, Cousinou, Covey, Crane, Cunha,
  Ilha, Dai, Damsted, Darling, Darrington, Davidson, Davies, Dawson, De, de~la
  Macorra, De~Lee, Queiroz, Machado, de~la Torre, Dell'Agli, Bourboux,
  Diamond-Stanic, Dillon, Donor, Drory, Duckworth, Dwelly, Ebelke,
  Eftekharzadeh, Eigenbrot, Elsworth, Eracleous, Erfanianfar, Escoffier, Fan,
  Farr, Fernandez-Trincado, Feuillet, Finoguenov, Fofie, Fraser-McKelvie,
  Frinchaboy, Fromenteau, Fu, Galbany, Garcia, Garcia-Hernandez, Oehmichen, Ge,
  Maia, Geisler, Gelfand, Goddy, Goff, Gonzalez-Perez, Grabowski, Green, Grier,
  Guo, Guy, Harding, Hasselquist, Hawken, Hayes, Hearty, Hekker, Hogg,
  Holtzman, Hou, Hsieh, Huber, Hunt, Chitham, Imig, Jaber, Angel, Johnson,
  Jones, Jonsson, Jullo, Kim, Kinemuchi, Kirkpatrick, Kite, Klaene, Kneib,
  Kollmeier, Kong, Kounkel, Krishnarao, Lacerna, Lan, Lane, Law, Leung, Lewis,
  Li, Lian, Lin, Long, Longa-Pena, Lundgren, Lyke, Mackereth, MacLeod,
  Majewski, Manchado, Maraston, Martini, Masseron, Masters, Mathur, McDermid,
  Merloni, Merrifield, Meszaros, Miglio, Minniti, Minsley, Miyaji, Mohammad,
  Mosser, Mueller, Muna, Munoz-Gutierrez, Myers, Nadathur, Nair, Nascimento,
  Nevin, Newman, Nidever, Nitschelm, Noterdaeme, O'Connell, Olmstead, Oravetz,
  Oravetz, Osorio, Pace, Padilla, Palanque-Delabrouille, Palicio, Pan, Pan,
  Parker, Paviot, Peirani, Ramrez, Penny, Percival, Perez-Fournon,
  Perez-Rafols, Petitjean, Pieri, Pinsonneault, Poovelil, Povick, Prakash,
  Price-Whelan, Raddick, Raichoor, Ray, Rembold, Rezaie, Riffel, Riffel, Rix,
  Robin, Roman-Lopes, Roman-Zuniga, Rose, Ross, Rossi, Rowlands, Rubin,
  Salvato, Sanchez, Sanchez-Menguiano, Sanchez-Gallego, Sayres, Schaefer,
  Schiavon, Schimoia, Schlafly, Schlegel, Schneider, Schultheis, Schwope, Seo,
  Serenelli, Shafieloo, Shamsi, Shao, Shen, Shetrone, Shirley, Aguirre, Simon,
  Skrutskie, Slosar, Smethurst, Sobeck, Sodi, Souto, Stark, Stassun, Steinmetz,
  Stello, Stermer, Storchi-Bergmann, Streblyanska, Stringfellow, Stutz, Suarez,
  Sun, Taghizadeh-Popp, Talbot, Tayar, Thakar, Theriault, Thomas, Thomas,
  Tinker, Tojeiro, Toledo, Tremonti, Troup, Tuttle, Unda-Sanzana, Valentini,
  Vargas-Gonzalez, Vargas-Magana, Vazquez-Mata, Vivek, Wake, Wang, Weaver,
  Weijmans, Wild, Wilson, Wilson, Wolthuis, Wood-Vasey, Yan, Yang, Yeche,
  Zamora, Zarrouk, Zasowski, Zhang, Zhao, Zhao, Zheng, Zheng, Zhu, \&
  Zou}]{Ahumada2020TheSpectra}
Ahumada, R., Prieto, C.~A., Almeida, A., {et~al.} 2020, arXiv

\bibitem[{Barnes(1988)}]{Barnes1988EncountersGalaxies}
Barnes, J.~E. 1988, \apj, 331, 699

\bibitem[{Bellazzini {et~al.}(2006)Bellazzini, Newberg, Correnti, Ferraro, \&
  Monaco}]{Bellazzini2006DetectionStream}
Bellazzini, M., Newberg, H.~J., Correnti, M., Ferraro, F.~R., \& Monaco, L.
  2006, \aap, 457, 21

\bibitem[{Belokurov {et~al.}(2018)Belokurov, Erkal, Evans, Koposov, \&
  Deason}]{Belokurov2018Co-formationHalo}
Belokurov, V., Erkal, D., Evans, N.~W., Koposov, S.~E., \& Deason, A.~J. 2018,
  \mnras, 478, 611

\bibitem[{Breddels \& Veljanoski(2018)}]{Breddels2018}
Breddels, M.~A. \& Veljanoski, J. 2018, \aap, 618, 13

\bibitem[{Chou {et~al.}(2007)Chou, Majewski, Cunha, Smith, Patterson,
  Martinez‐Delgado, Law, Crane, Munoz, Garcia~Lopez, Geisler, \&
  Skrutskie}]{Chou2007AStream}
Chou, M.-Y., Majewski, S.~R., Cunha, K., {et~al.} 2007, \apj, 670, 346

\bibitem[{Dalton {et~al.}(2012)Dalton, Trager, Abrams, Carter, Bonifacio,
  Aguerri, MacIntosh, Evans, Lewis, Navarro, Agocs, Dee, Rousset, Tosh,
  Middleton, Pragt, Terrett, Brock, Benn, Verheijen, Cano~Infantes, Bevil,
  Steele, Mottram, Bates, Gribbin, Rey, Rodriguez, Delgado, Guinouard, Walton,
  Irwin, Jagourel, Stuik, Gerlofsma, Roelfsma, Skillen, Ridings, Balcells,
  Daban, Gouvret, Venema, \& Girard}]{Dalton2012WEAVE:Telescope}
Dalton, G., Trager, S.~C., Abrams, D.~C., {et~al.} 2012, in Ground-based and
  Airborne Instrumentation for Astronomy IV, ed. I.~S. McLean, S.~K. Ramsay, \&
  H.~Takami, Vol. 8446 (International Society for Optics and Photonics), 84460P

\bibitem[{de~Jong {et~al.}(2012)de~Jong, Bellido-Tirado, Chiappini, Depagne,
  Haynes, Johl, Schnurr, Schwope, Walcher, Dionies, Haynes, Kelz, Kitaura,
  Lamer, Minchev, M{\"{u}}ller, Nuza, Olaya, Piffl, Popow, Steinmetz, Ural,
  Williams, Winkler, Wisotzki, Ansorge, Banerji, Gonzalez~Solares, Irwin,
  Kennicutt, King, McMahon, Koposov, Parry, Sun, Walton, Finger, Iwert, Krumpe,
  Lizon, Vincenzo, Amans, Bonifacio, Cohen, Francois, Jagourel, Mignot, Royer,
  Sartoretti, Bender, Grupp, Hess, Lang-Bardl, Muschielok, B{\"{o}}hringer,
  Boller, Bongiorno, Brusa, Dwelly, Merloni, Nandra, Salvato, Pragt, Navarro,
  Gerlofsma, Roelfsema, Dalton, Middleton, Tosh, Boeche, Caffau, Christlieb,
  Grebel, Hansen, Koch, Ludwig, Quirrenbach, Sbordone, Seifert, Thimm,
  Trifonov, Helmi, Trager, Feltzing, Korn, \&
  Boland}]{deJong20124MOST:Telescope}
de~Jong, R.~S., Bellido-Tirado, O., Chiappini, C., {et~al.} 2012, in
  Ground-based and Airborne Instrumentation for Astronomy IV, ed. I.~S. McLean,
  S.~K. Ramsay, \& H.~Takami, Vol. 8446 (International Society for Optics and
  Photonics), 84460T

\bibitem[{{DESI Collaboration} {et~al.}(2016){DESI Collaboration}, Aghamousa,
  Aguilar, Ahlen, Alam, Allen, Prieto, Annis, Bailey, Balland, Ballester,
  Baltay, Beaufore, Bebek, Beers, Bell, Bernal, Besuner, Beutler, Blake,
  Bleuler, Blomqvist, Blum, Bolton, Briceno, Brooks, Brownstein, Buckley-Geer,
  Burden, Burtin, Busca, Cahn, Cai, Cardiel-Sas, Carlberg, Carton, Casas,
  Castander, Cervantes-Cota, Claybaugh, Close, Coker, Cole, Comparat, Cooper,
  Cousinou, Crocce, Cuby, Cunningham, Davis, Dawson, de~la Macorra, De~Vicente,
  Delubac, Derwent, Dey, Dhungana, Ding, Doel, Duan, Ealet, Edelstein,
  Eftekharzadeh, Eisenstein, Elliott, Escoffier, Evatt, Fagrelius, Fan,
  Fanning, Farahi, Farihi, Favole, Feng, Fernandez, Findlay, Finkbeiner,
  Fitzpatrick, Flaugher, Flender, Font-Ribera, Forero-Romero, Fosalba, Frenk,
  Fumagalli, Gaensicke, Gallo, Garcia-Bellido, Gaztanaga, Fusillo, Gerard,
  Gershkovich, Giannantonio, Gillet, Gonzalez-de Rivera, Gonzalez-Perez, Gott,
  Graur, Gutierrez, Guy, Habib, Heetderks, Heetderks, Heitmann, Hellwing,
  Herrera, Ho, Holland, Honscheid, Huff, Hutchinson, Huterer, Hwang, Laguna,
  Ishikawa, Jacobs, Jeffrey, Jelinsky, Jennings, Jiang, Jimenez, Johnson,
  Joyce, Jullo, Juneau, Kama, Karcher, Karkar, Kehoe, Kennamer, Kent,
  Kilbinger, Kim, Kirkby, Kisner, Kitanidis, Kneib, Koposov, Kovacs, Koyama,
  Kremin, Kron, Kronig, Kueter-Young, Lacey, Lafever, Lahav, Lambert, Lampton,
  Landriau, Lang, Lauer, Goff, Guillou, Van~Suu, Lee, Lee, Leitner, Lesser,
  Levi, L'Huillier, Li, Liang, Lin, Linder, Loebman, Luki{\'{c}}, Ma, MacCrann,
  Magneville, Makarem, Manera, Manser, Marshall, Martini, Massey, Matheson,
  McCauley, McDonald, McGreer, Meisner, Metcalfe, Miller, Miquel, Moustakas,
  Myers, Naik, Newman, Nichol, Nicola, da~Costa, Nie, Niz, Norberg, Nord,
  Norman, Nugent, O'Brien, Oh, Olsen, Padilla, Padmanabhan, Padmanabhan,
  Palanque-Delabrouille, Palmese, Pappalardo, P{\^{a}}ris, Park, Patej,
  Peacock, Peiris, Peng, Percival, Perruchot, Pieri, Pogge, Pollack, Poppett,
  Prada, Prakash, Probst, Rabinowitz, Raichoor, Ree, Refregier, Regal, Reid,
  Reil, Rezaie, Rockosi, Roe, Ronayette, Roodman, Ross, Ross, Rossi, Rozo,
  Ruhlmann-Kleider, Rykoff, Sabiu, Samushia, Sanchez, Sanchez, Schlegel,
  Schneider, Schubnell, Secroun, Seljak, Seo, Serrano, Shafieloo, Shan,
  Sharples, Sholl, Shourt, Silber, Silva, Sirk, Slosar, Smith, Smoot, Som,
  Song, Sprayberry, Staten, Stefanik, Tarle, Tie, Tinker, Tojeiro, Valdes,
  Valenzuela, Valluri, Vargas-Magana, Verde, Walker, Wang, Wang, Weaver,
  Weaverdyck, Wechsler, Weinberg, White, Yang, Yeche, Zhang, Zhao, Zheng, Zhou,
  Zhou, Zhu, Zou, \& Zu}]{DESICollaboration2016TheDesign}
{DESI Collaboration}, Aghamousa, A., Aguilar, J., {et~al.} 2016, arXiv, 000

\bibitem[{Dohm-Palmer {et~al.}(2001)Dohm-Palmer, Helmi, Morrison, Mateo,
  Olszewski, Harding, Freeman, Norris, \&
  Shectman}]{Dohm-Palmer2001MAPPINGBODY}
Dohm-Palmer, R.~C., Helmi, A., Morrison, H., {et~al.} 2001, \apj, 555, L37

\bibitem[{Eneev {et~al.}(1973)Eneev, Kozlov, \&
  Sunyaev}]{Eneev1973TidalGalaxies}
Eneev, T.~M., Kozlov, N.~N., \& Sunyaev, R.~A. 1973, \aap, 22, 41

\bibitem[{{Gaia Collaboration} {et~al.}(2018){Gaia Collaboration}, Brown,
  Vallenari, Prusti, de~Bruijne, Babusiaux, \&
  Bailer-Jones}]{GaiaCollaboration2018brown}
{Gaia Collaboration}, Brown, A. G.~A., Vallenari, A., {et~al.} 2018, \aap, 616,
  21

\bibitem[{{Gaia Collaboration} {et~al.}(2016){Gaia Collaboration}, Prusti, J~de
  Bruijne, A~Brown, Vallenari, Babusiaux, L~Bailer-Jones, Bastian, Biermann,
  Evans, Eyer, Jansen, Jordi, Klioner, Lammers, Lindegren, Luri, Mignard,
  Milligan, Panem, Poinsignon, Pourbaix, Randich, Sarri, Sartoretti, Siddiqui,
  Soubiran, Valette, van Leeuwen, Walton, Aerts, Arenou, Cropper, Drimmel,
  H{\o}g, Katz, Lattanzi, Grebel, Holland, Huc, Passot, Bramante, Cacciari,
  Casta{\~{n}}eda, Chaoul, Cheek, De~Angeli, Fabricius, Guerra,
  Hern{\'{a}}ndez, Jean-Antoine-Piccolo, Masana, Messineo, Mowlavi,
  Nienartowicz, Ord{\'{o}}{\~{n}}ez-Blanco, Panuzzo, Portell, Richards,
  Barache, Barata, Barbier, Barblan, Baroni, Barrado~Navascu{\'{e}}s, Barros,
  Barstow, Becciani, Bellazzini, Bellei, Bello~Garc{\'{i}}a, Belokurov,
  Bendjoya, Berihuete, Bianchi, Bienaym{\'{e}}, Billebaud, Blagorodnova,
  Blanco-Cuaresma, Boch, Bombrun, Borrachero, Bouquillon, Bourda, Bouy,
  Bragaglia, Breddels, Brouillet, Br{\"{u}}semeister, Bucciarelli, Budnik,
  Burgess, Burgon, Burlacu, Busonero, Buzzi, Caffau, Cambras, Campbell,
  Cancelliere, Cantat-Gaudin, Carlucci, Carrasco, Castellani, Charlot, Charnas,
  Charvet, Chassat, Chiavassa, Clotet, Cocozza, Collins, Collins, Costigan,
  Crifo, G~Cross, Crosta, Crowley, Dafonte, Damerdji, Dapergolas, David, David,
  De~Cat, de~Felice, de~Laverny, De~Luise, De~March, de~Martino, de~Souza,
  Debosscher, del Pozo, Delbo, Delgado, Delgado, di~Marco, Di~Matteo, Diakite,
  Distefano, Dolding, Dos~Anjos, Drazinos, Dur{\'{a}}n, Dzigan, Ecale,
  Edvardsson, Enke, Erdmann, Escolar, Espina, Evans, Eynard~Bontemps, Fabre,
  Fabrizio, Faigler, Falc{\~{a}}o, Farr{\`{a}}s~Casas, Faye, Federici,
  Fedorets, Fern{\'{a}}ndez-Hern{\'{a}}ndez, Fernique, Fienga, Figueras,
  Filippi, Findeisen, Fonti, Fouesneau, Fraile, Fraser, Fuchs, Furnell, Gai,
  Galleti, Galluccio, Garabato, Garc{\'{i}}a-Sedano, Gar{\'{e}}, Garofalo,
  Garralda, Gavras, Gerssen, Geyer, Gilmore, Girona, Giuffrida, Gomes,
  Gonz{\'{a}}lez-Marcos, Gonz{\'{a}}lez-N{\'{u}}{\~{n}}ez,
  Gonz{\'{a}}lez-Vidal, Granvik, Guerrier, Guillout, Guiraud, G{\'{u}}rpide,
  Guti{\'{e}}rrez-S{\'{a}}nchez, Guy, Haigron, Hatzidimitriou, Haywood, Heiter,
  Helmi, Hobbs, Hofmann, Holl, Holland, S~Hunt, Hypki, Koubsky, Kowalczyk,
  Krone-Martins, Kudryashova, Kull, Bachchan, Lacoste-Seris, Lanza, Lavigne,
  Le~Poncin-Lafitte, Lebreton, Lebzelter, Leccia, Leclerc, Lecoeur-Taibi,
  Lemaitre, Lenhardt, Leroux, Liao, Licata, P~Lindstr{\o}m, Lister, Livanou,
  Lobel, L{\"{o}}ffler, L{\'{o}}pez, Lopez-Lozano, Lorenz, Loureiro, MacDonald,
  Magalh{\~{a}}es~Fernandes, Managau, Mann, Mantelet, Marchal, Pagani, Pagano,
  Pailler, Palacin, Palaversa, Parsons, Paulsen, Pecoraro, Pedrosa,
  Pentik{\"{a}}inen, Pereira, Pichon, Piersimoni, Pineau, Plachy, Plum,
  Poujoulet, Pr{\v{s}}a, Pulone, Ragaini, Rago, Rambaux, Ramos-Lerate, Ranalli,
  Rauw, Read, Regibo, Renk, Reyl{\'{e}}, Ribeiro, Rimoldini, Ripepi, Riva,
  Rixon, Roelens, Romero-G{\'{o}}mez, Rowell, Royer, Rudolph, Ruiz-Dern,
  Sadowski, Sagrist{\`{a}}~Sell{\'{e}}s, Sahlmann, Salgado, Salguero, Sarasso,
  Savietto, Schnorhk, Schultheis, Sciacca, Segol, Segovia, Segransan, Serpell,
  Shih, Smareglia, Smart, Smith, Solano, Solitro, Sordo, Soria~Nieto, Souchay,
  Spagna, Spoto, Stampa, Steele, Steidelm{\"{u}}ller, Stephenson, Stoev, Suess,
  S{\"{u}}veges, Surdej, Szabados, Szegedi-Elek, Tapiador, Taris, Tauran,
  Taylor, Teixeira, Terrett, Tingley, Trager, Turon, Ulla, Utrilla, Valentini,
  van Elteren, Van~Hemelryck, van Leeuwen, Varadi, Vecchiato, Veljanoski, Via,
  Vicente, Vogt, Voss, Votruba, Voutsinas, Walmsley, Weiler, Weingrill, Werner,
  Wevers, Whitehead, Wyrzykowski, Yoldas, {\v{Z}}erjal, Zucker, Zurbach,
  Zwitter, Alecu, Allen, Allende~Prieto, Amorim, Anglada-Escud{\'{e}},
  Arsenijevic, Azaz, Balm, Beck, Bernstein, Bigot, Bijaoui, Blasco, Bonfigli,
  Bono, Boudreault, Bressan, Brown, Brunet, Bunclark, Buonanno, Butkevich,
  Carret, Carrion, Chemin, Ch{\'{e}}reau, Corcione, Darmigny, de~Boer,
  de~Teodoro, de~Zeeuw, Delle~Luche, Domingues, Dubath, Fodor, Fr{\'{e}}zouls,
  Fries, Fustes, Fyfe, Gallardo, Gallegos, Gardiol, Gebran, Gomboc,
  G{\'{o}}mez, Grux, Gueguen, Heyrovsky, Hoar, Iannicola, Isasi~Parache,
  Janotto, Joliet, Jonckheere, Keil, Kim, Klagyivik, Klar, Knude, Kochukhov,
  Kolka, Kos, Kutka, Lainey, LeBouquin, Liu, Loreggia, Makarov, Marseille,
  Martayan, Martinez-Rubi, Massart, Meynadier, Mignot, Munari, Nguyen,
  Nordlander, Ocvirk, Olias~Sanz, Ortiz, Osorio, Oszkiewicz, Ouzounis, Palmer,
  Park, Pasquato, Peltzer, Peralta, P{\'{e}}turaud, Pieniluoma, \&
  Pigozzi}]{GaiaCollaboration2016TheMission}
{Gaia Collaboration}, Prusti, T., J~de Bruijne, J.~H., {et~al.} 2016, \aap,
  595, A1

\bibitem[{Hayes {et~al.}(2020)Hayes, Majewski, Hasselquist, Anguiano, Shetrone,
  Law, Schiavon, Cunha, Smith, Beaton, Price-Whelan, Prieto, Battaglia,
  Bizyaev, Brownstein, Cohen, Frinchaboy, Garc{\'{i}}a, Roman-Lopes, Sobeck, \&
  Stringfellow}]{Hayes2020MetallicityAPOGEE}
Hayes, C.~R., Majewski, S.~R., Hasselquist, S., {et~al.} 2020, \apj, 889, 63

\bibitem[{Helmi {et~al.}(2018)Helmi, Babusiaux, Koppelman, Massari, Veljanoski,
  \& Brown}]{Helmi2018}
Helmi, A., Babusiaux, C., Koppelman, H.~H., {et~al.} 2018, \nat, 563, 85

\bibitem[{Helmi \& de~Zeeuw(2000)}]{helmi2000}
Helmi, A. \& de~Zeeuw, P.~T. 2000, \mnras, 319, 657

\bibitem[{Helmi \& White(1999)}]{Helmi1999a}
Helmi, A. \& White, S. D.~M. 1999, \mnras, 307, 495

\bibitem[{Ho {et~al.}(2015)Ho, Kudritzki, Kewley, Zahid, Dopita, Bresolin, \&
  Rupke}]{Ho2015MetallicityMetals}
Ho, I.-T., Kudritzki, R.-P., Kewley, L.~J., {et~al.} 2015, \mnras, 448, 2030

\bibitem[{Hunter(2007)}]{Hunter2007Matplotlib:Environment}
Hunter, J.~D. 2007, Computing in Science {\&} Engineering, 9, 90

\bibitem[{Ibata {et~al.}(1994)Ibata, Gilmore, \& Irwin}]{Ibata1994ASagittarius}
Ibata, R.~A., Gilmore, G., \& Irwin, M.~J. 1994, \nat, 370, 194

\bibitem[{Jean-Baptiste {et~al.}(2017)Jean-Baptiste, Di~Matteo, Haywood,
  G{\'{o}}mez, Montuori, Combes, \& Semelin}]{Jean-Baptiste2017OnTale}
Jean-Baptiste, I., Di~Matteo, P., Haywood, M., {et~al.} 2017, \aap, 604, A106

\bibitem[{Katz {et~al.}(2019)Katz, Sartoretti, Cropper, Panuzzo, Seabroke,
  Viala, Benson, Blomme, Jasniewicz, Jean-Antoine, Huckle, Smith, Baker, Crifo,
  Damerdji, David, Dolding, Fr{\'{e}}mat, Gosset, Guerrier, Guy, Haigron,
  Jan{\ss}en, Marchal, Plum, Soubiran, Th{\'{e}}venin, Ajaj, Allende~Prieto,
  Babusiaux, Boudreault, Chemin, Delle~Luche, Fabre, Gueguen, Hambly, Lasne,
  Meynadier, Pailler, Panem, Royer, Tauran, Zurbach, Zwitter, Arenou, Bossini,
  Gerssen, G{\'{o}}mez, Lemaitre, Leclerc, Morel, Munari, Turon, Vallenari, \&
  Erjal}]{Katz2019GaiaVelocities}
Katz, D., Sartoretti, P., Cropper, M., {et~al.} 2019, \aap, 622, 19

\bibitem[{Kirby {et~al.}(2011)Kirby, Lanfranchi, Simon, Cohen, \&
  Guhathakurta}]{Kirby2011MULTI-ELEMENT}
Kirby, E.~N., Lanfranchi, G.~A., Simon, J.~D., Cohen, J.~G., \& Guhathakurta,
  P. 2011, \apj, 727, 78

\bibitem[{Kluyver {et~al.}(2016)Kluyver, Ragan-Kelley, P{\'{e}}rez, Granger,
  Bussonnier, Frederic, Kelley, Hamrick, Grout, Corlay, Ivanov, Avila, Abdalla,
  Willing, \& Development~Team}]{Kluyver2016JupyterWorkflows}
Kluyver, T., Ragan-Kelley, B., P{\'{e}}rez, F., {et~al.} 2016, {Jupyter
  Notebooks-a publishing format for reproducible computational workflows} (IOS
  Press)

\bibitem[{Kollmeier {et~al.}(2017)Kollmeier, Zasowski, Rix, Johns, Anderson,
  Drory, Johnson, Pogge, Bird, Blanc, Brownstein, Crane, De~Lee, Klaene,
  Kreckel, MacDonald, Merloni, Ness, O'Brien, Sanchez-Gallego, Sayres, Shen,
  Thakar, Tkachenko, Aerts, Blanton, Eisenstein, Holtzman, Maoz, Nandra,
  Rockosi, Weinberg, Bovy, Casey, Chaname, Clerc, Conroy, Eracleous,
  G{\"{a}}nsicke, Hekker, Horne, Kauffmann, McQuinn, Pellegrini, Schinnerer,
  Schlafly, Schwope, Seibert, Teske, \& van
  Saders}]{Kollmeier2017SDSS-V:Spectroscopy}
Kollmeier, J.~A., Zasowski, G., Rix, H.-W., {et~al.} 2017, arXiv

\bibitem[{Koppelman {et~al.}(2019)Koppelman, Helmi, Massari, Price-Whelan, \&
  Starkenburg}]{Koppelman2019}
Koppelman, H.~H., Helmi, A., Massari, D., Price-Whelan, A.~M., \& Starkenburg,
  T.~K. 2019, \aap, 631, L9

\bibitem[{Koppelman {et~al.}(2018)Koppelman, Helmi, \&
  Veljanoski}]{Koppelman2018}
Koppelman, H.~H., Helmi, A., \& Veljanoski, J. 2018, \apjl, 860, L11

\bibitem[{Lindegren(2018)}]{Lindegren2018b}
Lindegren, L. 2018, {Re-normalising the astrometric chi-square in Gaia DR2},
  Tech. rep., Lund Observatory

\bibitem[{Mackereth {et~al.}(2019)Mackereth, Schiavon, Pfeffer, Hayes, Bovy,
  Anguiano, Prieto, Hasselquist, Holtzman, Johnson, Majewski, O'connell,
  Shetrone, Tissera, \& Fer{\'{n}}andez-Trincado}]{Mackereth2019TheSimulations}
Mackereth, J.~T., Schiavon, R.~P., Pfeffer, J., {et~al.} 2019, \mnras, 482,
  3426

\bibitem[{Martinez-Delgado {et~al.}(2004)Martinez-Delgado, Gomez-Flechoso,
  Aparicio, \& Carrera}]{Martinez-Delgado2004TracingGalaxy}
Martinez-Delgado, D., Gomez-Flechoso, M.~A., Aparicio, A., \& Carrera, R. 2004,
  \apj, 601, 242

\bibitem[{Massari {et~al.}(2019)Massari, Koppelman, \&
  Helmi}]{Massari2019OriginWay}
Massari, D., Koppelman, H.~H., \& Helmi, A. 2019, \aap, 630, L4

\bibitem[{Matsuno {et~al.}(2019)Matsuno, Aoki, \& Suda}]{Matsuno2019}
Matsuno, T., Aoki, W., \& Suda, T. 2019, \apjl, 874, L35

\bibitem[{McMillan(2017)}]{Mcmillan2017}
McMillan, P.~J. 2017, \mnras, 94, 76

\bibitem[{McMillan \& Binney(2008)}]{McMillan2008}
McMillan, P.~J. \& Binney, J.~J. 2008, \mnras, 390, 429

\bibitem[{Morrison {et~al.}(1990)Morrison, Flynn, \& Freeman}]{Morrison1990}
Morrison, H.~L., Flynn, C., \& Freeman, K.~C. 1990, \aj, 100, 1191

\bibitem[{Myeong {et~al.}(2019)Myeong, Vasiliev, Iorio, Evans, \&
  Belokurov}]{Myeong2019EvidenceHalo}
Myeong, G.~C., Vasiliev, E., Iorio, G., Evans, N.~W., \& Belokurov, V. 2019,
  \mnras, 000, 1235

\bibitem[{Quinn(1984)}]{Quinn1984OnGalaxies}
Quinn, P.~J. 1984, \apj, 279, 596

\bibitem[{Quinn \& Goodman(1986)}]{Quinn1986SINKINGSYSTEMS}
Quinn, P.~J. \& Goodman, J. 1986, \apj, 309, 472

\bibitem[{Sch{\"{o}}nrich {et~al.}(2010)Sch{\"{o}}nrich, Binney, \&
  Dehnen}]{Schonrich2010}
Sch{\"{o}}nrich, R., Binney, J., \& Dehnen, W. 2010, \mnras, 403, 1829

\bibitem[{Thompson(2014)}]{Thompson2014PyGadgetReader:Python}
Thompson, R. 2014, Astrophysics Source Code Library, record ascl:1411.001

\bibitem[{Toomre \& Toomre(1972)}]{Toomre1972GalaxyTails}
Toomre, A. \& Toomre, J. 1972, \apj, 178, 623

\bibitem[{Tormen {et~al.}(1998)Tormen, Diaferio, \&
  Syer}]{Tormen1998SurvivalHaloes}
Tormen, G., Diaferio, A., \& Syer, D. 1998, \mnras, 299, 728

\bibitem[{Van Den~Bosch {et~al.}(1999)Van Den~Bosch, Lewis, Lake, \&
  Stadel1}]{VanDenBosch1999SUBSTRUCTUREFRICTION}
Van Den~Bosch, F.~C., Lewis, G.~F., Lake, G., \& Stadel1, J. 1999, \apj, 515,
  50

\bibitem[{Van Der~Walt {et~al.}(2011)Van Der~Walt, Colbert, \&
  Varoquaux}]{VanDerWalt2011TheComputation}
Van Der~Walt, S., Colbert, S.~C., \& Varoquaux, G. 2011, Computing in Science
  and Engineering, 13, 22

\bibitem[{Vasiliev(2019)}]{Vasiliev2019}
Vasiliev, E. 2019, \mnras, 482, 1525

\bibitem[{Villalobos \& Helmi(2008)}]{Villalobos2008}
Villalobos, A. \& Helmi, A. 2008, \mnras, 391, 1806

\bibitem[{Villalobos \& Helmi(2009)}]{Villalobos2009SimulationsDiscs}
Villalobos, A. \& Helmi, A. 2009, \mnras, 399, 166

\end{thebibliography}

\appendix

\section{Observational datasets used} 

\subsection{The RVS sample}
\label{app:RVS}

In Fig.~\ref{fig:velocity}, we use the subset with full phase-space information from {\it Gaia} DR2 as comparison and to interpret the simulations. The dataset, known as the RVS sample, contains $7~224~631$ stars. We only consider stars with ${\tt parallax\_over\_error}>5$, ${\tt parallax} > 1.0~{\rm mas}$, and ${\tt RUWE} < 1.4$ \citep[see][]{Lindegren2018b}, which leaves $2~784~095$ sources. 

For this set of stars, we calculate cylindrical space velocities using the transformations implemented in {\tt vaex} \citep{Breddels2018}. We place the Sun at $X = -8.2~{\rm kpc}$, correct for the motion of the local standard of rest (LSR) using $v_{\rm LSR}=232.8~{\rm km/s}$ \citep[both based on][]{Mcmillan2017}, and for the solar motion with respect to the LSR using $(U_\odot,V_\odot,W_\odot) = (11.1, 12.24, 7.25)~{\rm km/s}$ \citep{Schonrich2010}. Finally, we use a kinematic selection to filter stars from the disc. All the stars with $|{\bf V-V}_{LSR}|<210~{\rm km/s}$ are removed. The resulting set of stars comprises $5~371$ sources that we label as halo stars.

\subsection{Known halo structures and their abundances}
The structures that are shown in Fig.~\ref{fig:APOGEE} originate from \cite{Koppelman2019}. In summary, the selection criteria for stars belonging to Gaia-Enceladus and Sequoia are defined as
\begin{enumerate}
    \item Gaia-Enceladus: $-1.5 < En< -1.1$ in units of $[10^5~{\rm km^2/s^{2}}]$ and $-0.20 < circ < 0.13$;
    \item Sequoia: $-1.35 < En < -1.0$ in units of $[10^5~{\rm km^2/s^{2}}]$ and $-0.65 < circ < -0.4$.
\end{enumerate}
Here, $En$ is the total energy, using the \cite{Mcmillan2017} potential and $circ$ (i.e. circularity) is defined as $    circ = \frac{L_z}{L_z^{circ}}$,
where $L_z^{circ}$ is the angular momentum of a circular orbit with the same energy as $L_z$. The circularity parameter ranges from $-1$ to $1$ and indicates how circular an orbit is.

The abundances that are being shown in the figure come from a cross-match of {\it Gaia} DR2 with APOGEE DR16 \citep{Ahumada2020TheSpectra}. Following the work of \cite{Hayes2020MetallicityAPOGEE}, we filter stars with {\tt STARFLAG} bitmask values of 0, 3, or 4 and {\tt ASPCAPFLAG} bitmask values of 10 or 23. Furthermore, we only show stars with ${\tt VERR}<0.2$, ${\tt SNR}>70$, and ${\tt TEFF}>3700$.

\section{Computation of the orbital parameters for the simulations}
\label{app:orbpar}
To calculate orbital properties (i.e. energy and eccentricity) we fit a static potential to the simulation snapshot at $t = 10$ Gyr after infall. In this snapshot, the debris of the satellite is sufficiently mixed for the whole system to be approximated by a smooth, rigid potential. 

We fit the potential with {\tt AGAMA} \citep{Vasiliev2019} using a {\it Multipole} expansion to approximate each component (dark matter and stellar) of both the host and satellite separately. For the flattened stellar particle distributions we set ${\tt gridsizeR=gridsizez}=50$, ${\tt mmax} = 5$, and ${\tt lmax} = 25$. We tested using a {\it CylSpline} for the flattened stellar systems but found the {\it Multipole} to reproduce the density profile better. The dark matter profile of the host is forced to be spherically symmetric, the other profiles are allowed to be triaxial. 

The energy of the particles is computed for this potential at this same time, $t = 10$ Gyr, and labelled $(E_f)$, as is the angular momentum $L_z$. The eccentricity is derived by integrating the stellar particles' orbits for 20 Gyr in this same potential. 


\end{document}